\documentclass[
reprint,
groupedaddress,
showpacs,
showkeys,
amsmath, amssymb,
aps,
prl,
]{revtex4-1}

\usepackage{xcolor}
\usepackage{graphicx}
\usepackage{dcolumn}
\usepackage{bm}
\usepackage[colorlinks=true, allcolors=blue]{hyperref}

\usepackage[separate-uncertainty=false]{siunitx}

\usepackage{xspace}
\newcommand{\na}{$^{23}$Na\xspace}

\newcommand{\ket}[1]{\vert#1\rangle}

\usepackage{xstring}
\newcommand*{\aref}[1]{%
	\IfBeginWith{#1}{eq:}{Eq.~\eqref{#1}}{}%
	\IfBeginWith{#1}{fig:}{Fig.~\ref{#1}}{}%
	\IfBeginWith{#1}{tab:}{Table~\ref{#1}}{}%
	\IfBeginWith{#1}{appendix:}{Appendix~\ref{#1}}{}%
	\IfBeginWith{#1}{sec:}{Section~\ref{#1}}{}%
}

\newcommand{\uu}{$\uparrow \uparrow$\xspace}
\newcommand{\ud}{$\uparrow \downarrow$\xspace}

\begin{document}

\title{Observation of magnetic solitons in two-component Bose-Einstein condensates}
\date{\today}

\author{A. Farolfi}
\author{D. Trypogeorgos}
\author{C. Mordini}
\author{G. Lamporesi}
\email[]{giacomo.lamporesi@ino.it}
\homepage[]{http://bec.science.unitn.it}
\author{G. Ferrari}
\affiliation{INO-CNR BEC Center and Dipartimento di Fisica, Universit\`a di Trento, 38123 Povo, Italy}
\affiliation{Trento Institute for Fundamental Physics and Applications, INFN, 38123 Povo, Italy}

\begin{abstract}
We experimentally investigate the dynamics of spin solitary waves (magnetic solitons) in a harmonically trapped, binary superfluid mixture.
We measure the in-situ density of each pseudospin component and their relative local phase via an interferometric technique we developed, and as such, fully characterise the magnetic solitons while they undergo oscillatory motion in the trap.
Magnetic solitons exhibit non-dispersive, dissipationless long-time dynamics.
By imprinting multiple magnetic solitons in our ultracold gas sample,  we engineer binary collisions between solitons of either same or opposite magnetisation and map out their trajectories. 

\end{abstract}

\keywords{}

\maketitle


Waves have the natural tendency to spread while propagating. In nonlinear media, this tendency can be counterbalanced through a self-focusing mechanism  creating localized and long-lived solitary waves, a.k.a. solitons. Their dissipationless nature makes them invaluable tools for technological applications and information transport \cite{Hermann96,Nakazawa00}. 
They play a fundamental role across science, classical and quantum alike, and have been observed in different physical systems, such as classical fluids, liquid He, plasmas, optical waveguides, polaritons, and ultracold atomic gases \cite{Ancilotto18,Lonngren83,Dauxois06,Amo11,Carr08,Anderson08}.
The latter can be widely manipulated to explore soliton behaviour, by altering the shape of the gas, the characteristic interactions among particles, and their energy dispersion~\cite{Burger99,Denschlag00,Anderson01,Strecker02,Khaykovich02,Eiermann04,Becker08, Shomroni09,Hamner11,Ku16}.

\begin{figure}[b]
    \centering
    \includegraphics[width=\columnwidth]{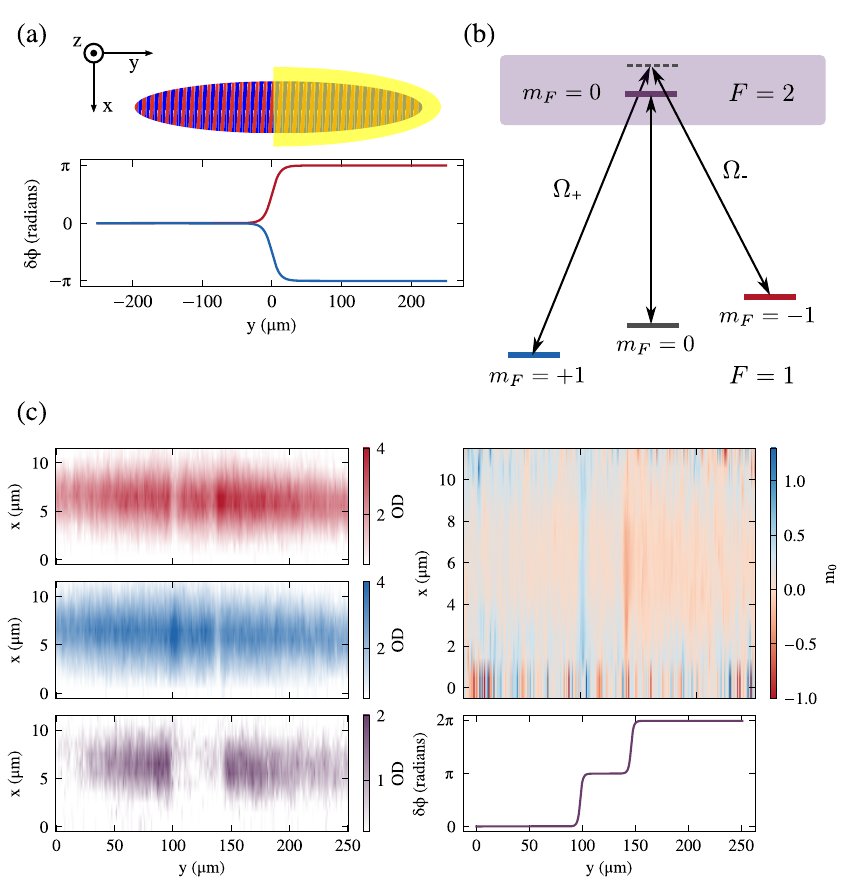}
    \caption{(a) A spin-selective optical potential generates a pair of MSs that travel in opposite directions along $y$ in our two-component elongated BEC.
    The overall imparted phase of $2\pi$ is dealt symmetrically on the two spin components.
    (b) $\Lambda$-coupling scheme showing all the hyperfine transitions that are used for preparation of the mixture and inducing an effective quadratic shift.
    (c) Full tomography of a pair of MSs 15\,ms after their creation. 
    Left column: Optical densities (OD) of $\ket{1,-1}$ (red) and $\ket{1,+1}$ (blue), and relative phase (purple). 
    Right column: The measured apparent magnetisation (top) is of the order of 0.5 and the expected relative-phase profile (bottom) shows two $\pi$ jumps at the soliton positions.}
    \label{fig:figure1}
\end{figure}

Two-component mixtures display an even richer excitation spectrum, 
showing new types of solitons. These solitons were long-sought in the liquid He community, but were never observed due to the absence of an experimental realisation of interpenetrable superfluids. However, mixtures of ultracold atomic gases can be used instead \cite{Ferrier-Barbut14,Bienaime16,Danaila16,Fava18,Kim19arxiv}.  A mixture can be perturbed from its ground state by creating either excitations in the total density, with an in-phase response of the two components, or excitations in the population imbalance (magnetisation), with an out-of-phase response. 
This implies the existence of both unmagnetised solitons, similar to those in a single component superfluid, and magnetised ones \cite{Danaila16}.
Among the latter, \textit{magnetic solitons} (MS) are denoted by a localised population imbalance in an otherwise balanced and symmetrically interacting mixture \cite{Qu16}.

Atomic mixtures (superpositions) of \na lowest-hyperfine-state atoms in the $\ket{F,m_\text{F}}=\ket{1,\pm 1}$  are fully miscible and not subject to buoyancy~\footnote{The intra- and intercomponent scattering lengths for \na are $a=54.54\, a_0$ and  $a_\pm=50.78\, a_0$ respectively~\cite{Knoop11}, and correspond to an interaction difference $\delta a=a-a_{\pm}=3.76\,a_0=0.07\,a$.}. 
The two ground-state components experience the same trapping potential, show the same spatial profile, and occupy the same volume \cite{Bienaime16,Fava18}. These are prerequisite conditions for the excitation and characterisation  of MSs \cite{Qu16}, which are fulfilled in our system \cite{Bienaime16}; however this is not the case in other atomic species, such as $^{87}$Rb \cite{Danaila16}.

Here, we create MSs via spin-sensitive phase imprinting.
We characterize them in-situ using a fully tomographic method with quasi-concurrent density and relative-phase measurements, that show a characteristic $\pi$ jump. 
The MSs perform oscillatory dynamics in a harmonically confined BEC that show only minimal dispersion and dissipation for times as long as 1\,sec.
In addition, we engineer collisions between MSs with same \uu and opposite \ud magnetisation  and monitor their behaviour close to the collision point.

\textit{Experiment.} 
All  experiments described here begin with a thermal cloud of \na atoms in a hybrid trap~\cite{Lamporesi13,Colzi16,Colzi18,Farolfi19} in the $\ket{1,-1}$ state, which we then transfer into an elongated crossed optical trap (\aref{fig:figure1}a). Further evaporative cooling leads to a BEC of typically $N\simeq2\times 10^6$ atoms with negligible thermal component, $T\simeq 250$\,nK.
The final trap frequencies are $\{f_y,\,f_\perp\} = \{8.7(1.2),\,585(2)\}$\,Hz giving axial and transverse Thomas-Fermi radii $R_y\simeq250\,\mu$m and $R_\perp\simeq3.7\,\mu$m, respectively. 
A uniform magnetic field is applied along the $z$-axis with a Larmor frequency of 182.3(1)\,kHz.
Atoms are transferred into a mixture of $\ket{1, \pm 1}$ via a two-photon microwave radiation with an effective Rabi frequency of 268(2)\,Hz using an adiabatic rapid passage technique; an initially large detuning of $\approx 4$\,kHz is gradually reduced to zero in $60$\,ms (see \aref{fig:figure1}b). 
After the two-photon coupling is switched off, a dressing radiation with Rabi frequency 2.27(5)\,kHz is turned on, 20\,kHz blue-detuned from $\ket{1,0} \rightarrow \ket{2, 0}$, that creates an effective quadratic shift and stabilizes the mixture against spin-relaxation~\cite{Gerbier06,Bienaime16,JimenezGarcia19}.

We produce MSs by applying a step-like, purely vectorial, optical dipole potential to the right half of the BEC (\aref{fig:figure1}a)~\footnote{We set the scalar part of the potential equal to zero by an appropriate choice of the laser frequency $\lambda=589.557$\,nm, which is detuned to the blue of the $3^2S_{1/2} \to 3^2P_{1/2}$ \na transition and to the red of the $3^2S_{1/2} \to 3^2P_{3/2}$ transition~\cite{Grimm00}}.
The light is circularly polarized in order to maximize the vector term of the light shift and, since the atomic states have opposite angular momentum, the phase imprinted on the $\ket{1, \pm 1}$ states is opposite by construction, $\phi_+=-\phi_-=\phi/2$ (\aref{fig:figure1}a). 
Using a pulsetime of $\tau = 70$\,$\mu$s we imprint a phase of $\pm \pi$ onto the $\ket{1,\pm 1}$ states. The amount of imprinted phase is independently calibrated (see Supplemental Materials).
The phase imprint pulse does not introduce additional spin or density excitations since $\tau < h/ng \ll h/n\delta g$, where $n$ is the total atomic density, $g$ is the intracomponent interaction constant, $\delta g $ the difference between intra- and intercomponent interactions, and $h$ is Planck's constant. 
The light beam is elliptical with an aspect ratio of 5:1 and its masked intensity goes from 10\% to 90\% over 2\,$\mu$m $\simeq 3\xi_s$, where $\xi_s=\hbar/\sqrt{2mn\, \delta g}$ is the spin healing length and $m$ the atomic mass. 
This produces two MSs since the total magnetisation of the system is conserved. 
They move in opposite directions, have opposite magnetisation, and are robust against transverse instabilities. 

The stability of a soliton depends on the ratio between its transverse extension and its thickness \cite{Brand02,Komineas03}. 
If the latter is much smaller than the former, the soliton resembles a fragile thin membrane that decays into vortical structures through snaking instability~\cite{Anderson01,Shomroni09,Becker13,Mateo14,Ku16}.
Assuming the same stability criterion valid for density solitons with the relevant quantities replaced by their spin equivalents, we expect stability for $R_\perp \lesssim 6\xi_s$.
In our system $R_\perp \simeq 5\xi_s$ and the MSs show stable one-dimensional dynamics.

Time-of-flight measurements do not reveal any phase dislocations ~\cite{Shomroni09,Becker13,Ku14,Donadello14,Tylutki16}--- associated with the formation of vortices or vortex rings --- in either component even after 1\,sec from the creation of MSs (see Supplemental Materials).
Density dynamics of the BEC are still three-dimensional, with phase coherence along the whole sample since $\mu/\hbar \omega_\perp=9.3$, where $\mu$ is the chemical potential of the BEC.

\begin{figure}[t]
    \centering
    \includegraphics[width=\columnwidth]{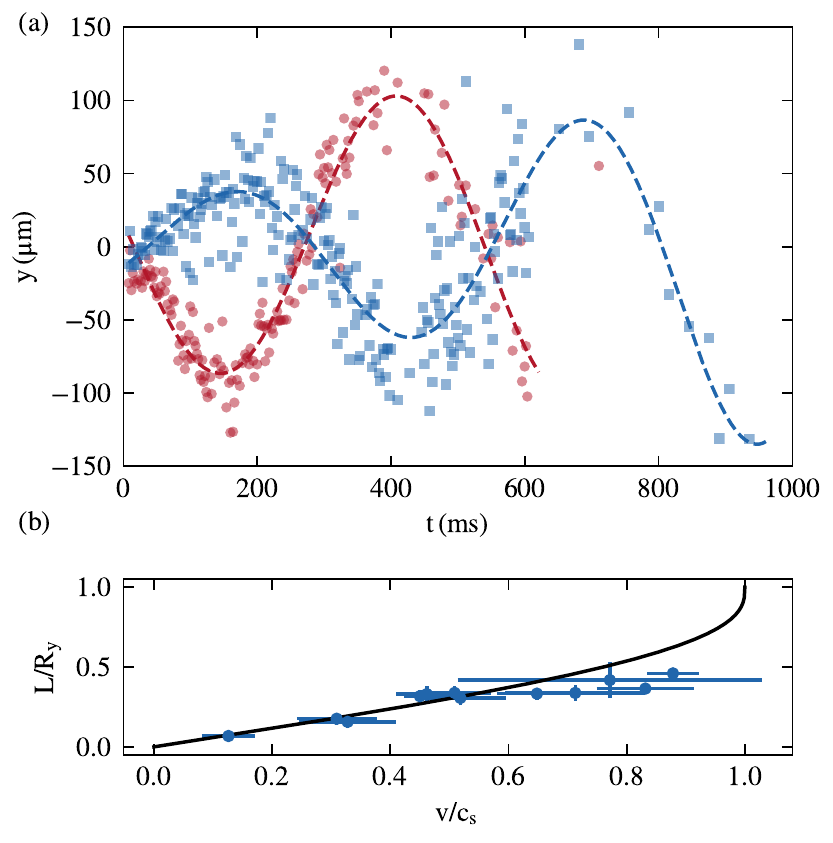}
    \caption{(a) Two MSs of opposite magnetisation oscillate out-of-phase in the trap with a period of 550\,ms. 
    The BEC is centered at $y=0$.
    Finite temperature effects lead to the slight oscillation amplitude increase and decay of magnetisation (see Supplemental Material). Since the magnetisation of one of the MSs is slightly smaller, it goes below our detection threshold sooner.
    (b) The oscillation amplitude $L/R_y$ of a single MS increases monotonically with its initial velocity $v$, in agreement with the theoretical prediction with no free parameters~\cite{Qu16}.
    }
    \label{fig:figure2}
\end{figure}

\textit{Dynamics.} 
We detect the density profiles of the two components and their relative phase throughout the sample (\aref{fig:figure1}c).  
We measure density in-situ by separately transferring 14\% of the atoms from $\ket{1,+1}$ and $\ket{1,-1}$ to the $\ket{2,0}$ state, and then imaging them using the $F=2 \rightarrow F'=3$ transition. 
Where one spin component shows a local density dip, the other one has a peak and vice versa. 
Together they comprise MSs with positive or negative magnetisation, while the total density is unperturbed.
The two transfers happen 600\,$\mu$s apart in which time the soliton has traversed a distance of $\approx$600\,nm which is smaller than our optical resolution of roughly 2\,$\mu$m.
After 2.5\,ms we acquire an image of the relative phase of the two components.
Figure~\ref{fig:figure1}c shows a full tomographic snapshot of two MSs 16\,ms after their creation at the center of the trap.
They have an opposite apparent magnetisation $|m_0|=|n_+-n_-|/n=0.5$ and have travelled to $\pm 20$\,$\mu$m with a velocity $v\simeq$1.2\,mm/s; the relative phase shows two discontinuities at these positions.

Figure~\ref{fig:figure2}a shows the MS long-time dynamics in the harmonic trap for up to 1 second since their formation. 
They undergo multiple oscillations with no discernible dispersion. 
The oscillation amplitude is $\approx 0.4R_y$ and the period $4.7(1)T_y$, where  $T_y$ the axial harmonic oscillator period. 
Throughout the oscillatory dynamics we observe a slight oscillation amplitude increase and decay of magnetisation, a dissipation likely induced by the residual thermal fraction, while our limited imaging resolution does not allow to appreciate an evolution of the width of the MS.

Varying the BEC atom number and the intensity gradient of the phase imprint light we launch MSs with different $v/c_s$, where $c_s=\sqrt{n\,\delta g/2m}$ is the spin sound velocity.
Figure~\ref{fig:figure2}b shows the dependence of the amplitude of oscillation $L/R_y$ on the peak velocity $v/c_s$ and is in good agreement with theory~\cite{Qu16}.

\begin{figure}[t]
    \centering
    \includegraphics[width=\columnwidth]{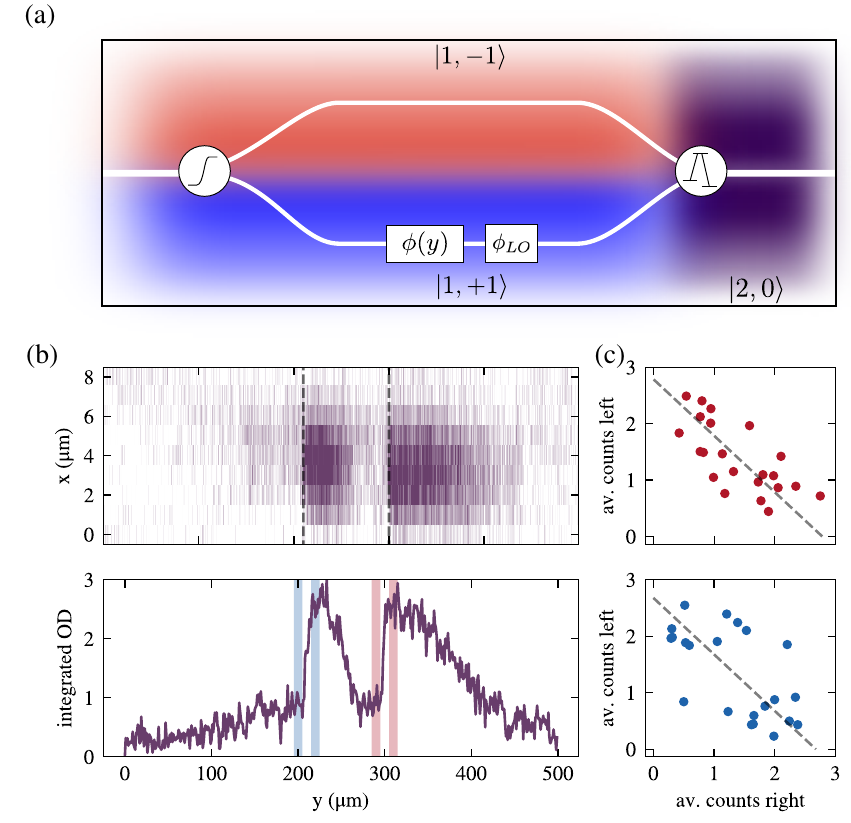}
    \caption{(a) The input state of the interferometer is an equal superposition of the two components (red-blue).
    All phases are applied to one arm without loss of generality.
    The two components are projected onto the readout state (purple) using a bichromatic microwave pulse.
    (b) Typical data showing the OD in the readout state (top), which is a direct measurement of the phase difference of the two components.
    The jump in $\phi(y)$ is proportional to the ratio of the average counts on either side of a MS (shaded regions).
    (c) Average counts in the left versus right regions across each of the two MSs.
 The phase difference of the left- and right-projected populations is independent of $\phi_{LO}$. The anticorrelated behaviour is consistent with a $\pi$ phase jump for both MSs within our statistical uncertainty.
    }
    \label{fig:figure3}
\end{figure}

\textit{Interferometer.}
We measure the relative phase of the two components using a generalised Ramsey technique.
The input state of the interferometer $\ket{1,-1} + \exp{\left[i(\phi(y) + \phi_{LO})\right]}\ket{1,+1}$ is an equal superposition of the ground hyperfine levels, where, without loss of generality, the $\ket{1,+1}$ state carries all the relative phase; we ignore any global phase associated with the unitary evolution of the Hamiltonian under Larmor precession.
The term $\phi(y)$ corresponds to the phase-imprint pulse which acts locally on the BEC and initially imprints a bipartite left/right phase which is later redistributed along the BEC due to the relative motion of the MSs.
The phase of the local oscillator $\phi_{LO}$ is the phase difference of the two microwave fields that form the interferometer and act globally on the BEC (see \aref{fig:figure3}a).
A bichromatic microwave pulse, resonant with the $\ket{1,-1} \to \ket{2,0}$ and $\ket{1,+1} \to \ket{2,0}$ transitions with respective Rabi frequencies $\Omega_-=\Omega_+=\Omega$ (\aref{fig:figure3}a), acts as a beamsplitter that projects the ground-excited-state superposition onto the population of $\ket{2, 0}$.
We then image the full spatial distribution of $P_{\ket{2,0}} = \left[1+\cos(\phi(y)+\phi_{LO})\right]\sin^2{\Omega t}$, where the populations of $\ket{1,\pm 1}$ are normalised to unity; the relative local phase of the two ground-state components is mapped onto the population of $\ket{2, 0}$.
The duration of the microwave pulse is 20\,$\mu$s and the two arms of the interferometer are balanced with $\Omega/2\pi = 1.12(1)$\,kHz (see Supplemental Material).

We are interested in how $\phi(y)$ changes across a MS. 
For any random $\phi_{LO}$, a $\pi$ jump in $\phi(y)$ results in the anti-correlated output of the interferometer when comparing the counts on the two sides of each MS. 
However effects such as the residual spin-dependent curvature of the potential and the excitation of long-wavelength spin waves~\cite{Hamner13}~\footnote{Note for instance the difference between Fig.~\ref{fig:figure1}c lower panel and Fig.~\ref{fig:figure3}b top panel which, despite being taken in similar experimental settings, show a qualitative different pattern.
From single magnetisation images of atomic clouds with a much larger aspect ratio 100:1, we estimate the energy curvature to be 450\,Hz/mm$^2$, or 28\,Hz for our $R_y = 250$\,$\mu$m BEC.}, adversely affects the output of the interferometer.
We circumvent these by restricting the phase measurement to
the regions adjacent to the position of the MSs (see shaded regions in \aref{fig:figure3}b). 
Figure~\ref{fig:figure3}c shows the anti-correlation in the interferometer output channels which appears as a distribution with $-1$ slope when plotted against each other.

\begin{figure}[t]
    \centering
    \includegraphics[width=\columnwidth]{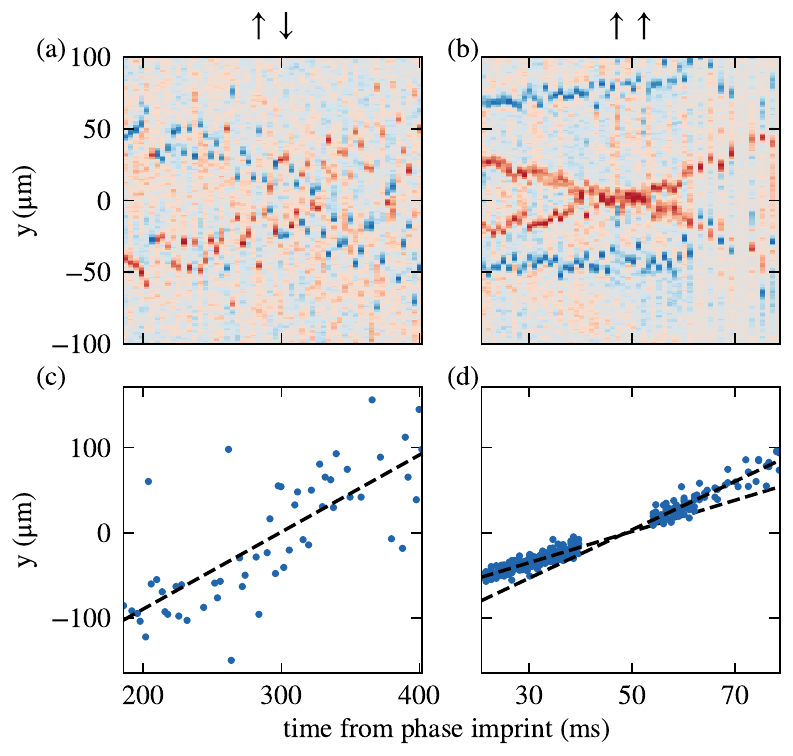}
    \caption{Collisions of MSs with opposite (left column) or same (right column) magnetisations.
    (a-b) Time evolution of the MSs in the reference frame of the center-of-mass of the solitons.
    (c-d) Relative position of the MSs close to the collision point at 300\,ms  for \ud collisions (c) and 45\,ms for \uu (d).
    For these \uu collisions ($m_0=0.86$), the MSs dissipate energy during the collision as evident by the different velocities (slopes) before and after.
    The missing points in (d) are due to not being able to distinguish same magnetisation MSs when they are very close together.
    The variance of the data is larger in the left column than in the right one since the collision happens at much later times.
    }
    \label{fig:figure4}
\end{figure}

\textit{Collisions.} 
We engineer \uu collisions by phase imprinting two pairs of MSs in our BEC, $\pm 50$\,$\mu$m from the center, using a different optical mask.
We ignore the two that travel towards the edges of the BEC and focus on the \uu pair that collides at the center 45\,ms later (\aref{fig:figure4}b).

Since the size of the MSs is smaller than our optical resolution, we 
infer their magnetisation from their velocity, instead.
The relative velocity of the two \uu MSs increases with respect to its pre-collision value from $1.83(8)$\,mm/s to $2.8(1)$\,mm/s (\aref{fig:figure4}d), implying a change in $m_0$ from 0.86 to 0.62.
Such a dissipative behaviour is in direct contrast with the expected solitonic interactions, that are normally dissipationless. Our observations can be naively explained by considering that for low-magnetisation \uu collisions, the total magnetisation at the collision is approximately the linear sum of the two MSs. However for $|m_0| > 0.5$ this would lead to magnetisation larger than 1, that is unphysical. The system responds to this apparent impasse by introducing dissipation.

Collisions of MSs with opposite magnetisation \ud happen naturally in our system when the two MSs reach the centre of the trap half an oscillation later (Fig.\ref{fig:figure2}a).
Figure~\ref{fig:figure4}a shows their trajectories in the reference frame where the center-of-mass of the solitons is fixed.  
In \ud collisions MSs go through each other, but our signal-to-noise ratio does not allow us to conclude whether there is a change in their relative velocity (\aref{fig:figure4}c).
The noise in \ud collisions is larger than the one in \uu collisions since the former happen 300\,ms instead of  45\,ms after the MSs creation.

\textit{Conclusions.} 
We produced and characterised various aspects of MSs using the tomographic techniques described above.
Our MSs are stable, nondispersive, relatively long-lived, and their dynamic behaviour is in good agreement with theory \cite{Qu16}.
Collisions of \uu solitons of large magnetisation show violation of the solitonic property of nondissipative interaction.
This dissipative behaviour is reminiscent of light bullets: non-linear, stable structures that appear in dispersive optical media and lose energy when they collide~\cite{Milian19}. Our observations may trigger further studies on soliton interaction mechanisms, including the role of finite temperature effects.

With a suitable low-magnetic-field-noise environment~\cite{Farolfi19,trypogeorgos18}, our techniques can be readily extended towards investigating the physics of MSs in the presence of coherent coupling between the two components~\cite{Qu17,Su13,Shamailov18}.
Note that this work does not correspond to the limit of zero coupling between the two components. The solitons supported by the coupled and uncoupled systems have a distinct topological character~\cite{sanz2019interaction,Qu17}, since the phase across a coherently-coupled MS cannot be continuously unwound to $\pm\pi$.

While finalising this manuscript we became aware of similar work on magnetic excitations~\cite{chai2019}.

\begin{acknowledgements}
The authors are grateful to S. Stringari, F. Dalfovo, L. Pitaevskii, A. Recati, C. Qu, A. Gallem\'i,  M. Lewenstein and L. Tarruell for stimulating discussions.
We acknowledge funding from the project NAQUAS of QuantERA ERA-NET Cofund in Quantum Technologies (Grant Agreement N. 731473) implemented within the European Union's Horizon 2020 Programme, from Provincia Autonoma di Trento and from INFN-TIFPA under the project FISh.
\end{acknowledgements}

\bibliography{bibliography.bib}

\newpage
\clearpage
\pagebreak
\begin{widetext}
\begin{center}
\textbf{\large Supplemental Materials}
\end{center}
\end{widetext}

\setcounter{equation}{0}
\setcounter{figure}{0}
\setcounter{table}{0}
\setcounter{footnote}{0}
\setcounter{page}{1}
\makeatletter
\renewcommand{\theequation}{S\arabic{equation}}
\renewcommand{\thefigure}{S\arabic{figure}}
\renewcommand{\bibnumfmt}[1]{[S#1]}
\section{MS detection}
\begin{figure}[h]
    \centering
    \includegraphics[width=\columnwidth]{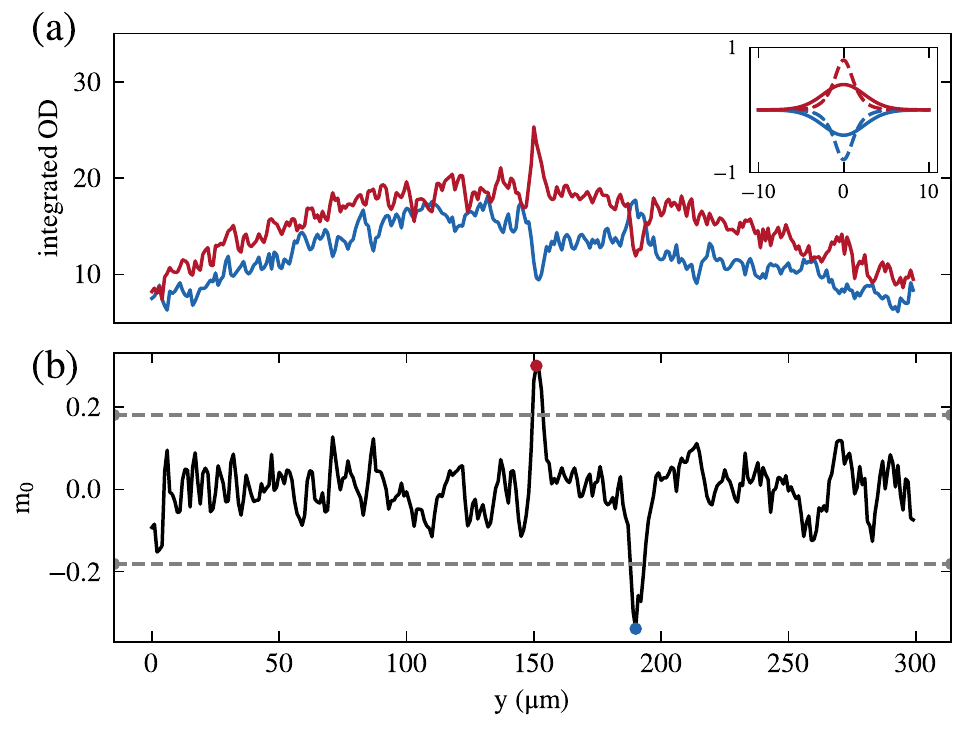}
    \caption{ (a) Profile of the integrated OD of the two components. The inset shows how the computed MS profile with $m_0=0.8$ (dashed) is broadened taking the finite optical resolution into account (solid). (b) Apparent magnetisation of the system along $y$. The points indicate the detected location of the MSs. Dashed lines show the threshold for peak detection.
    }
    \label{fig:S0}
\end{figure}

We identify the position of MSs in the sample by taking an in-situ image of the density distribution for each state of the mixture. The image is integrated along the $x$ axis to compute two 1D density distributions (\aref{fig:S0}a).

The inset shows the density profile of the two components of a MS moving with a velocity $v/c_s=0.6$. This corresponds to a MS with a width of $\approx \xi_s \approx 0.8\,\mu$m (dashed line) and a true magnetisation $m_0=0.8$. Due to our finite resolution $\approx 2\,\mu$m, the profile is broadened and the height is reduced (solid line).
We infer the true magnetisation from velocity measurements since $m_0=\sqrt{1-v^2/c_s^2}$.

The two 1D density distributions are used to compute the apparent magnetisation $m_0$. Due to the presence of long-wavelength spin excitations superimposed to the profile of the solitons, we apply a high-pass filter to the magnetization profile (\aref{fig:S0}b) before locating the position of the MSs. To avoid false positives, only the peaks with magnitude exceeding a given threshold (gray dashed lines) are considered as solitons (red and blue dots).

\section{Decay of the magnetic solitons}

The stability of a soliton is given by the ratio between its transverse extension and its thickness $R_\perp/\xi$~\cite{Brand02,Komineas03}. 
In the case of MSs the existence of dynamically stable solutions is expected for $R_\perp/\xi \lesssim 6$~\cite{Qu16}, while for larger values of this parameter the soliton resembles a fragile thin membrane that decays into vortical structures through the snaking instability~\cite{Denschlag00,Anderson01,Mateo14,Ku16}. 
Solitonic vortices and vortex rings are the most probable decay products, being the least energetic excitations supported by the system.

\begin{figure}[h]
    \centering
    \includegraphics[width=\columnwidth]{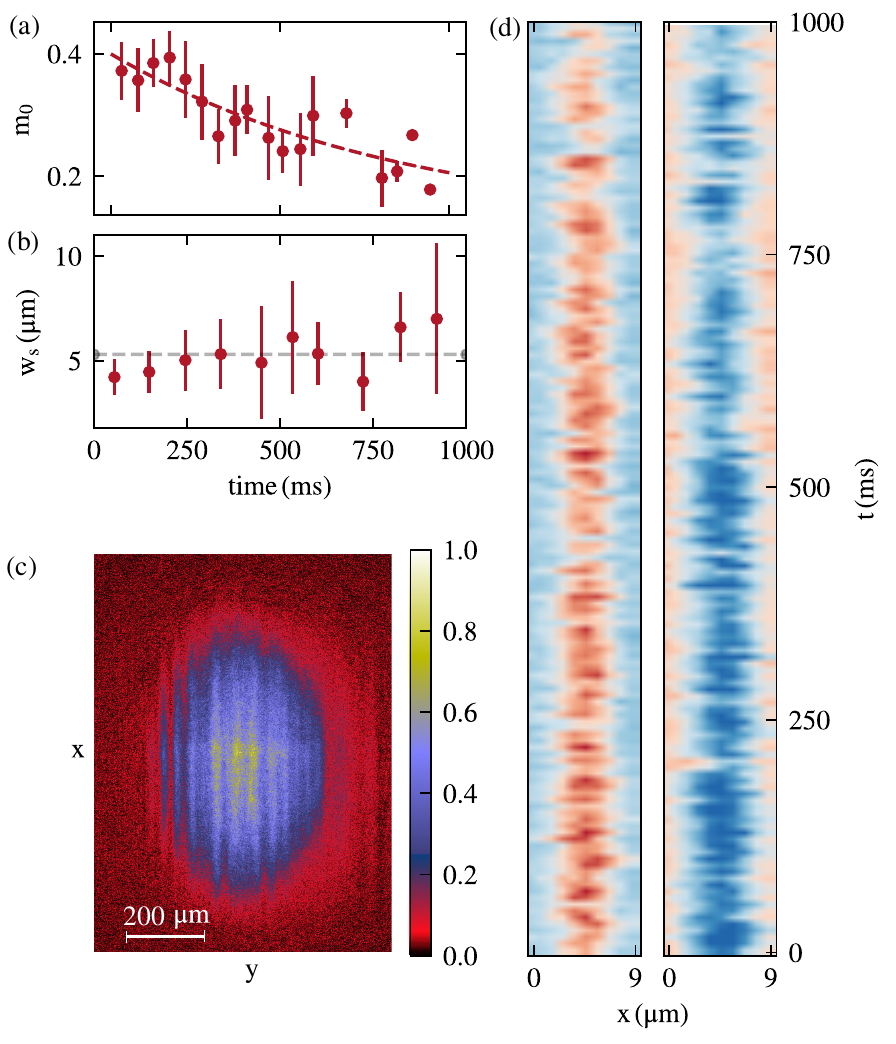}
    \caption{
    (a) MSs persist for about 1\,sec in our system before their peak magnetisation becomes comparable to our signal-to-noise ratio.
    (b) The apparent width of the MSs is constant throughout the lifetime of the MS.
    (c) Even after 700\,ms of in-trap dynamics, no phase distortions, characteristic of solitonic vortices, appear in images after a ballistic expansions of the gas for 30\,ms.
    (d) Profile of the magnetisation in the transverse $x$ direction, shown consecutively in time. The transverse shape of MSs seems to remain stable and shows no evidence of snaking instability.
    }
    \label{fig:S1}
\end{figure}

Since the transverse size of our BEC is of the order of critical value expected to allow relaxation of the MSs via snaking instability, it is important to verify the dynamical evolution of the MSs.
The lifetime of the MS is about 1\,sec, past which its magnetisation becomes comparable to the imaging noise in our system (\aref{fig:S1}a).
The apparent width (FWHM) of the MS remains constant to 5\,$\mu$m.
Its real width its of the order of $\xi_s$ which is smaller than our resolution.

We look for snaking-instability-like decay of the soliton in the transverse profile of the MS magnetisation. 
Curving (snaking) of the MS along the long axis of the BEC would manifest as an amplitude reduction along its projection on the transverse axis. 
In-situ imaging, which has a resolution of the order of the radial Thomas-Fermi radius does not provide any evidence of relaxation of the soliton into a vortex line or ring. 
In order to increase our sensitivity to vortical excitations we turn to imaging one single component after ballistic expansion but also in this case we never observe signatures possibly related to the presence of circulation in the sample such as localized dislocations along the depleted line of the soliton, or rotation of the soliton plane (see \aref{fig:S1}d).

The question about the fate of the MS when the binary mixture is not properly balanced arises naturally, as well as what happens in the finite-temperature regime. 
We do not systematically investigate unbalanced or finite-temperature mixtures since they are beyond the scope of the present work. 
On the other hand in the early stage of our measurements we notice that population imbalance often results in the formation of a single MS steadily centered at the center of the BEC 
and exhibiting a peak in the population of the most abundant component. 
With regard to the effect of temperature on the MS dynamics, we remark that we observe reliable MS oscillations of more than one period only after minimizing the temperature in the binary mixture prior to the phase imprint.

\section{Calibration of the phase imprint}

\begin{figure}[h]
    \centering
    \includegraphics[width=\columnwidth]{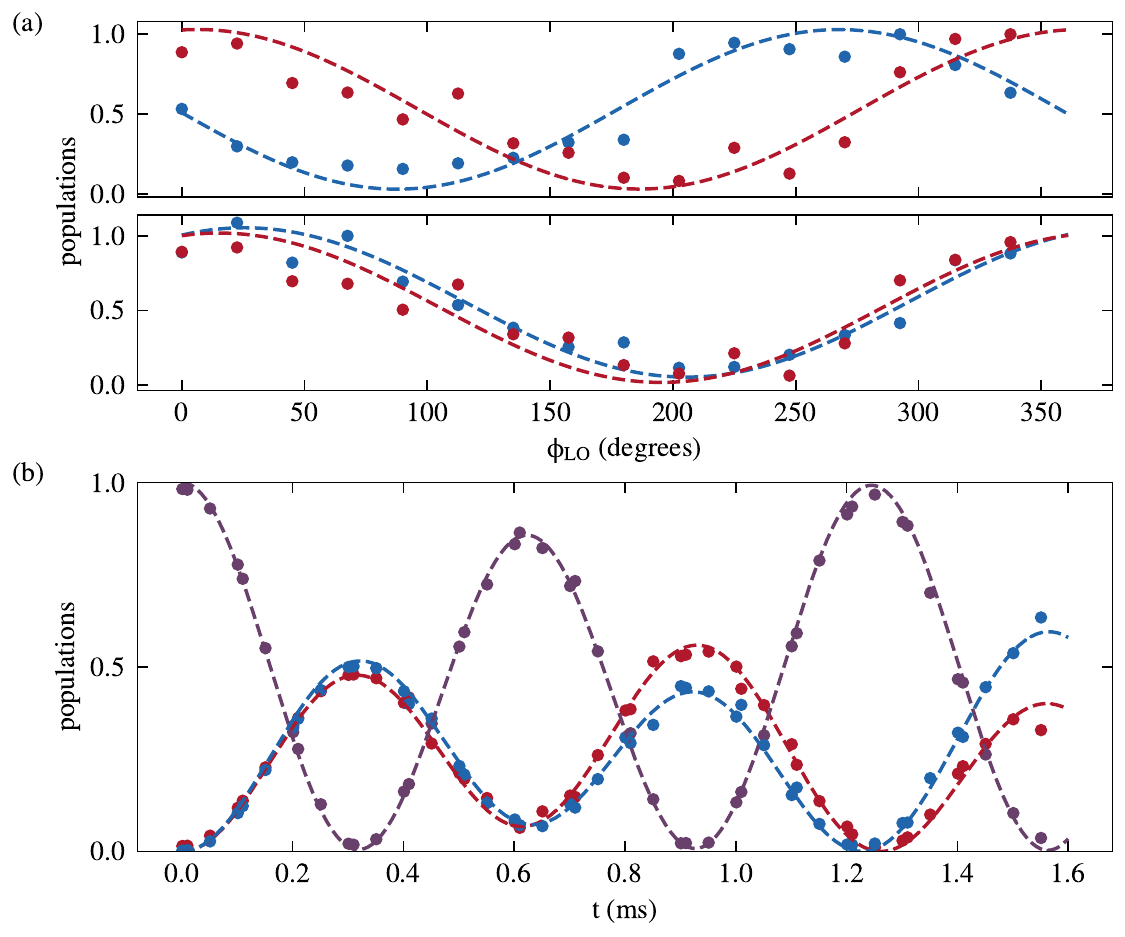}
    \caption{(a) Absolute calibration of the interferometer using the phase of the local oscillator to convert atom counts to radians.
    The measured population in $\ket{2,0}$ is plotted as a function of the set $\phi_{LO}$.
    The two bipartite splits have a relative phase of 2.6$\pi$ for a 90\,$\mu$s phase-imprint pulse (top), and 2$\pi$ for a 70\,$\mu$s (bottom).
    (b) Symmetric Rabi oscillations from $\ket{2, 0}$ to $\ket{1, \pm 1}$ show that the interferometer is balanced, with a Rabi frequency $\Omega/2\pi=1.12$\,kHz}
    \label{fig:S2}
\end{figure}

We calibrate the phase imprint pulse by measuring the relative phase imprinted on a balanced $\ket{1,\pm1}$ mixture. We vary the length of the pulse and measure the interferometer output in the illuminated and dark halves of the BEC. By scanning the local oscillator phase $\phi_{LO}$ for a given pulse time, we measure the relative phase imprinted. Figure \ref{fig:S2}a shows the outputs of the illuminated (top) and dark (bottom) half of the BEC for two different phase imprint pulses (red and blue).

A pre-requisite for this is for the ground-state population to be equally transferred to $\ket{2,0}$ so the Rabi frequencies of the two interferometer arms need to be the same and on resonance. 
In order to balance the interferometer, we use a reduced Rabi frequency (1.12(1)\,kHz) to avoid spectral crowding from saturation of the rf electronics.
Figure~\ref{fig:S2}b shows the populations of all three participating states. 
The system undergoes symmetric oscillations between $\ket{2,0}$ and $\ket{1,\pm1}$.
For this, we prepared the system in the $\ket{2,0}$ state and induced unitary dynamics with a resonant bichromatic microwave pulse.
The interferometric measurement allows us to readily access the imprinted relative phase $\phi$ and set its value to the desired amount $2\pi$.

\section{Spin sound excitations}
\begin{figure*}[t!]
    \centering
    \includegraphics[width=2\columnwidth]{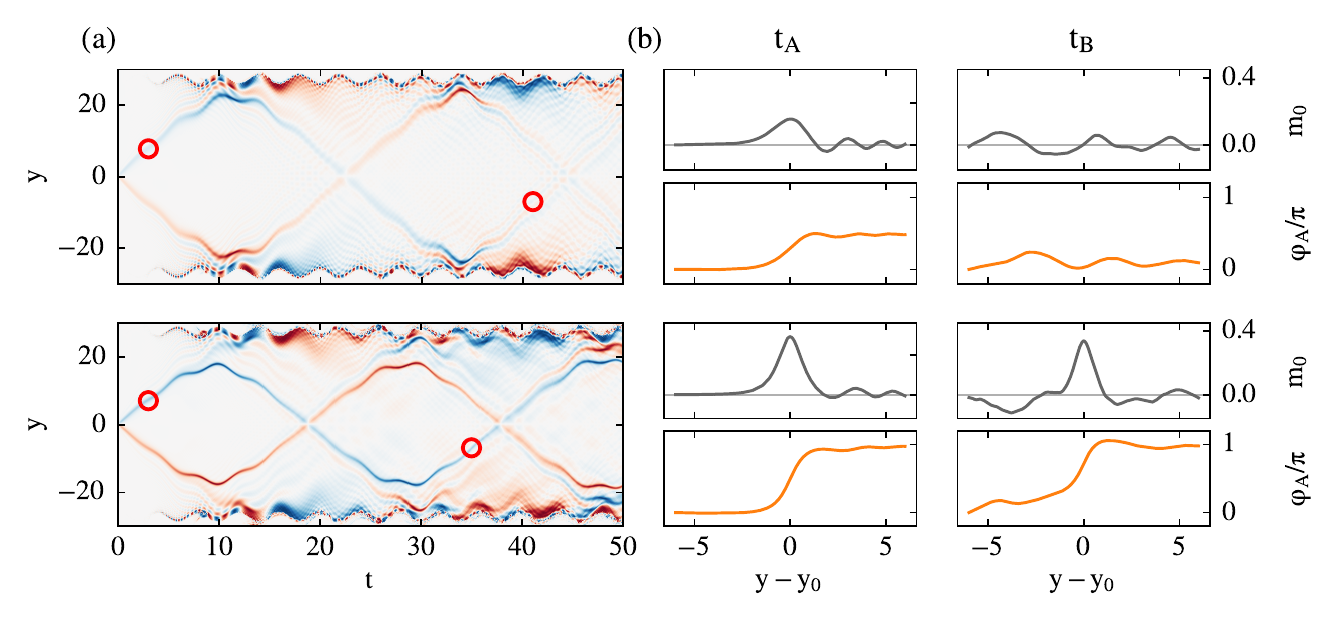}
    \caption{(a) NPSE simulation of the dynamics of two pairs of different kinds of magnetic excitations. In the upper (lower) row the initial phase imprint is $\pi$ (2$\pi$). The time and space coordinates are given in units of the harmonic oscillator frequency and length. (b) Magnetisation and jump in the relative phase across the magnetic excitation at time $t_A$ and $t_B$.}
    \label{fig:S3}
\end{figure*}

Two-component systems can sustain spin sound wave packets, travelling with a phase jump which is not locked to $\pi$ but is rather dependent on their group velocity. 
Due to the nonlinearity in the phononic dispersion relation, such excitations are expected to show a weak dispersive behaviour, increasing their width and losing contrast while propagating through the system. 
There is the possibility that these magnetic objects are excited together with MSs by the phase imprint procedure if the total phase difference imprinted by the optical potential is different from $2\pi$.

In order to gain insight on this spurious effect we simulate the generation of magnetic excitations after an imprint of a relative phase corresponding either to $\pi$ or 2$\pi$.
A two-component non-polynomial Schr\"odinger equation (NPSE) \cite{Salasnich2006} is suited to simulate the one-dimensional dynamics in our elongated system while taking into account the full three-dimensional scattering properties of the condensate in this density regime. The effective wave equation in the high-density regime ($a|\psi_{1,2}|^2 \gg 1$) reads
\begin{equation}
\begin{split}
    \label{eq:npse}
    i \hbar \partial_t \psi_1 = 
        &\left[ -\frac{\hbar^2}{2m}\partial_y^2 + \frac 1 2 m \omega_y^2 y^2 \right]\psi_1 + \\ 
        & \left[ \frac 3 2 \lambda \hbar \omega_y \sqrt{2a |\psi_1|^2 + 2a_{\pm} |\psi_2|^2} \right]\psi_1,
\end{split}
\end{equation}
for the first component, and symmetrically ($1 \leftrightarrow 2$) for the second component. Here $a$ and $a_{\pm}$ are respectively the intra- and intercomponent scattering lengths, $\omega_y = 2\pi f_y$, $\lambda = f_\perp / f_y$ is the aspect ratio of the system, and the equation is valid within the approximation $(a - a_{\pm}) / a = \delta g / g \ll 1$.

Figure~\ref{fig:S3} shows that in both cases two excitations with opposite magnetisations are created, but with different properties. For a total imprinted phase of $\pi$, spin sound wave packets are produced. They are dispersive and spread their width while they oscillate in the trap because of nonlinearity of the medium.
We show their shape at two different times, at $t_A$, shortly after their creation, and at $t_B$, nearly one oscillation period later.
Clearly the sound excitation (Fig.~\ref{fig:S3}a top) spreads over a larger region and changes its shape. 
When, instead, the total imprinted phase is $2\pi$ (Fig.~\ref{fig:S3}a bottom), a magnetic soliton is created, as clearly visible from its ability to preserve the initial shape and magnetisation in time.
The latter is much closer to what we observe experimentally, where the magnetization (Fig.~\ref{fig:S1}a) after one full oscillation is reduced to 75\% of the initial one.

\newpage


\end{document}